# Flickering Buoyant Diffusion Flames in Weakly Rotatory Flows

Tao Yang, Peng Zhang*

*Department of Mechanical Engineering, The Hong Kong Polytechnic University, Hung Hom, Kowloon, Hong Kong*

**Abstract**

Flickering buoyant diffusion methane flames in weakly rotatory flows were computationally and theoretically investigated. The prominent computational finding is that the flicker frequency nonlinearly increases with the rotational intensity number $R$ (up to 0.24), which measures the relative importance of the rotational speed compared with the methane jet speed. This finding is consistent with the previous experimental observations that flame flicker is enhanced by rotatory flows within a certain extent. Based on the vortex-dynamical understanding of flickering flames that the flame flicker is caused by the periodic shedding of buoyancy-induced toroidal vortices, we formulated a scaling theory for flickering buoyant diffusion flames in weakly rotatory flows. The theory predicts that, with respect to the flicker frequency $f_0$ at $R = 0$, the increase of the flicker frequency $f$ at nonzero $R$ obeys the scaling relation $f - f_0 \sim R^2$, which agrees very well with the present computational results. In physics, the externally rotatory flow enhances the radial pressure gradient around the flame, and the significant baroclinic effect $\nabla p \times \nabla \rho$ contributes an additional source for the growth of toroidal vortices so that their periodic shedding is faster.

---

* Corresponding author
  E-mail address: pengzhang.zhang@polyu.edu.hk
  Fax: (852)23654703, Tel: (852)27666664

**Nomenclature**

| | |
|---|---|
| $A$ | control mass |
| $\partial A$ | material contour of $A$ |
| $B(t)$ | time-dependent control volume enclosing the toroidal vortex |
| $C$ | constant threshold for vortex shedding |
| $C_j$ | constant prefactor relating to initial fuel jet |
| $C_r$ | constant prefactor relating to rotational flow |
| $C_{jr}$ | $C_j + C_r$ |
| $C_\theta$ | constant prefactor relating to azimuthal velocity within vortex core |
| $C_h$ | constant prefactor relating to vertical motion of toroidal vortex |
| $C_\rho$ | $\rho_\infty/\rho_f$ ranging $4-8$ |
| $D$ | diameter of fuel inlet, fixed at 10 mm |
| $f$ | flickering frequency |
| $\Delta f$ | $f(R) - f(R=0)$ |
| $g$ | gravitational constant |
| $h(t)$ | height of the upper boundary of $B(t)$ |
| $\bar{H}$ | time-averaged height of growing toroidal vortex, $\bar{H} = \hat{t}^{-1}\int_0^{\hat{t}} \hat{h}(t)dt$ |
| $Q$ | total heat release rate |
| $R$ | Rotational intensity number |
| $r_a$ | radius of vortex core |
| $r_c$ | radial position of vortex layer |
| $r(t)$ | width of the right boundary of $B(t)$ |
| $\bar{R}$ | time-averaged width of growing toroidal vortex, $\bar{R} = \hat{t}^{-1}\int_0^{\hat{t}} \hat{r}(t)dt$ |
| $t$ | time |
| $\hat{t}$ | dimensionless time, $t\sqrt{g/D}$ |
| $t^*$ | normalized time, $tf_0$ |
| $U_0$ | inlet velocity of fuel jet, fixed at 0.165m/s |
| $U$ | magnitude of inlet airflow, $\|\boldsymbol{U}\|$ |
| $u$ | velocity magnitude |
| $h$ | helicity density, $\boldsymbol{u}\cdot\boldsymbol{\omega}$ |
| $p$ | pressure |
| | |
| $\boldsymbol{a}$ | acceleration |
| $\boldsymbol{r}$ | unit radial direction vector |
| $\boldsymbol{s}$ | unit tangential vector along the contour $\partial A$ |
| $\boldsymbol{z}$ | unit normal vector |
| $\boldsymbol{u}$ | velocity vector |
| $\boldsymbol{U}$ | inlet velocity of airflow on the wind wall |
| $\boldsymbol{U}_\perp$ | normal component of $\boldsymbol{U}$ |
| $\boldsymbol{U}_\parallel$ | tangential component of $\boldsymbol{U}$ |
| | |
| $\widehat{\phantom{x}}$ | dimensionless quantity |

*Greek symbols*

| | |
|---|---|
| $\alpha$ | angle between the $U_\perp$ and $U_\parallel$, fixed at 45 degree |
| $\nu$ | viscosity |
| $\rho$ | density |
| $\tau$ | periodic time, $1/f$ |
| $\omega$ | vorticity magnitude |
| $\boldsymbol{\omega}$ | vorticity vector |

*Subscripts*

| | |
|---|---|
| 0 | quiescent environment ($R = 0$) |
| $x, y, z$ | components in Cartesian coordinates |
| $r, \theta, z$ | components in cylindrical coordinates |
| $f$ | flame |
| $\infty$ | environment condition at far field |

## 1. Introduction

In nature and in various domestic and industrial applications, diffusion flames could be unstable under buoyancy [1-12]. Some buoyant diffusion flames possess a well-known phenomenon, where the flame vibrates due to the periodic shedding of its upper portion, and the phenomenon is often referred to as "the flicker of luminous flames" [1, 2] or "fire puffing" [4, 9]. Many studies [4, 7, 10, 11] have substantiated that the flame flicker is a self-exciting flow oscillation and that the flickering frequency is proportional to $\sqrt{g/D}$, where $g$ is the gravitational constant and $D$ is the diameter of the flame burner (or fire pool). The experiment of Chen et al. [5] clearly illustrates that the shear layer around a diffusion flame rolls up due to the buoyancy-induced Kelvin–Helmholtz instability, evolves into a toroidal vortex, and eventually sheds off downstream, being synchronized with the flame flicker. Following the physics picture, Xia and Zhang [11] established a vortex-dynamical scaling theory to predict the flicker frequency of various pool and jet flames reported in the literature, and they confirmed and extended the correlation between the periodicity of toroidal vortices and the flickering buoyant diffusion flame.

The coupling interaction of multiple flickering flames can generate richer dynamical phenomena. In the system of dual flickering candle flames, Kitahata et al. [13] reported two distinct dynamical modes, namely the in-phase and anti-phase synchronizing modes at relatively small and large gap distances between flames, respectively. Similar phenomena can appear in not only candle flames [13-15] but also diffusion flames [16, 17] and pool flames [18]. Physically, the distinct modes are due to the interaction of the vortices generated around each flame [18], which is similar to the mechanism causing flow transition in the wake of a bluff body and forming the Kármán vortex street. The triple flickering flames [19-22] in various arrangements exhibit substantially more complex dynamical modes. From the perspective of vorticity reconnection and vortex-induced flow, Yang et al. [22] recently interpreted four typical dynamical modes (in-phase, flickering death, rotation, and partially in-phase) of triple flickering buoyant diffusion flames in equilateral triangle arrangement.

All the flickering flames mentioned above were produced in quiescent environment, and their interaction with externally forced flows is apparently of interest but insufficiently investigated. The stabilization of flames under rotatory flows is a long-lasting and practically relevant problem. In the past decades, many experiments [23-27] studied the influence of burner rotation on flame stabilization with the emphasis on the dynamical behaviors of buoyancy-induced flame oscillation. Gotoda together with coworkers [25-27] reported that the periodic flame flickering could retain at small rotational speeds but transition into low-dimensional deterministic chaos (flames exhibit spiral oscillation) at sufficiently large rotation speeds. Another experimental approach is to tangentially import airflow from the ambient into the central flame region [28-31]. This approach was usually used for producing a fire whirl, where whirling eddies of air, like a tornado flow, suck the ground fuel and form a slender fire downstream [32, 33]. Lei et al. [28] observed higher pulsation frequency of small-scale buoyant flames in rotatory flows compared with that in quiescent environment. In addition, Coenen et al. [29] found that the puffing instability of pool fires is suppressed under sufficiently strong rotating flows and a helical instability appears instead.

All the experimental studies mentioned above suggest that the flame oscillation frequency would be changed with the intensity of externally rotatory flows within a certain extent. However, a clear physics interpretation from the perspective of vortex dynamics is not available in the literature. In this study, we attempted to answer the following questions motivated from the previous experiments: how a flickering buoyant diffusion flame exhibits when it is subject to an externally rotating flow; how its flickering frequency varies with the rotating flow speed; what the vortex-dynamical origin of the observed frequency variation is. It should be emphasized that the present computational and theoretical investigations were limited to the situation of weakly rotatory flows so that the vortex breakdown does not appear in present problem although it certainly merits future studies.

The present paper is organized as follows. First, externally rotatory flows are generated by using four forced-ventilation walls and the rotational intensity is controlled by adjusting the ventilating velocity; flickering buoyant diffusion flames of methane gas are computationally produced and validated. Then, by comparing the evolution of toroidal vortices around flames under different rotational intensities, the effects of the rotatory flows on flickering fames are identified and analyzed. Finally, following Xia and Zhang's work [11], we established a vortex-dynamical scaling theory for flickering flames in weakly rotatory flows; the computational and theoretical results are compared and analyzed.

## 2. Computational Methodology and Validation

2.1 Computational Setup of Flickering Buoyant Diffusion Flames

In the present study, flickering buoyant diffusion flames were computationally produced using Fire Dynamics Simulator (FDS) [34], which is a widely used open-source code for solving the unsteady, three-dimensional, incompressible (variable-density) flow with combustion heat release. In the past decades, the code has been demonstrated to be a reliable computational platform for studying unsteady processes in fire-driven flows [18, 22, 35-40]. Based on FDS, our previous computational works successfully captured dynamical behaviors of flickering buoyant diffusion flames in quiescent environment and reproduced a variety of dynamical modes in the dual and triple flame systems [18, 22].

As shown in Fig. 1(a), the computational domain is a square column with $16D$ side and $24D$ height, where $D = 10$ mm is the fixed diameter of fuel inlet. A uniform structured mesh of $160 \times 160 \times 240$ is utilized as a balance of the computational cost and the grid independence, which was discussed in detail in the previous paper [18]. An impermeable, non-slip, and adiabatic solid-wall boundary is used at the bottom ground (grey area), while a central circle (yellow area) is open for the fuel inlet. The ambient air is the density of $\rho_\infty = 1.20$ kg/m$^3$ at the room temperature. The methane gas jet of the density of $\rho_F = 0.66$ kg/m$^3$ is ejected out at the uniform velocity of $U_0 = 0.165$ m/s to sustain a laminar diffusion flame with $Re = 100$ and $Fr = 0.28$, where $Re = U_0 D/\nu_F$ is the fuel-jet Reynold number ($v_F = 1.65 \times 10^{-5}$ m$^2$/s is the viscosity of methane gas) and $Fr = U_0^2/gD$ is the Froude number ($g = 9.8$ m/s$^2$ is the gravitational constant). The other sides of the computational domain are set as the non-solid exterior, through which gaseous products are allowed to flow in or out freely. To generate a rotatory flow environment for the flickering flame, forced-ventilation walls can be specified at four lateral sides of the computational domain for tangentially ejecting airflows into the central region. The generated rotatory flows will be illustrated and analyzed shortly in Sec. 2.2. In the present study, the fundamental conservation equations governing fluid dynamics are solved based on the explicit, second-order, kinetic-energy-conserving numerics and shown in details in the previous paper [18]. The turbulence modeling is not needed in the present laminar flows. The mixing-limited, infinitely fast reaction is used for modeling the present diffusion flames being far from extinction. The soot and radiation formation are not modeled in the present small-scale flames. The present computational validation of flickering buoyant diffusion flames will be expounded shortly in Sec. 2.3.

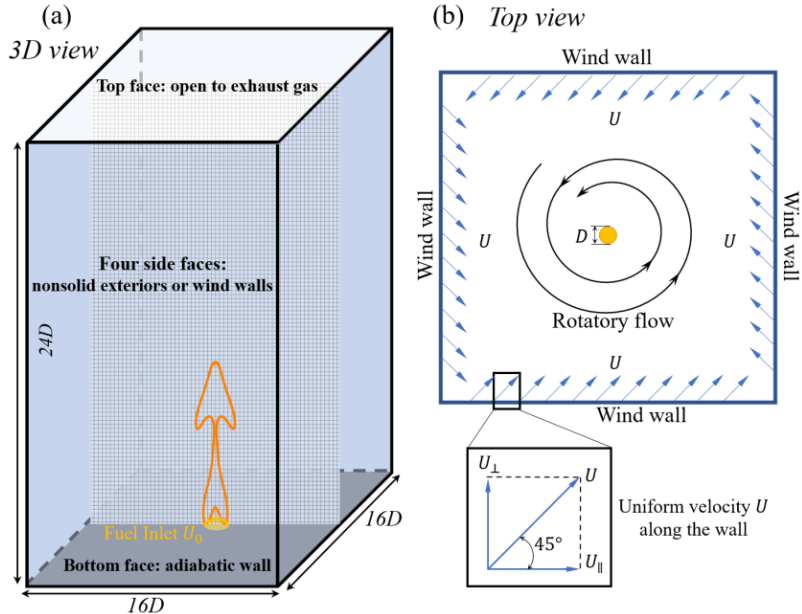

FIG. 1. (a) Schematic of the computational domain, mesh, and boundary conditions; (b) The rotatory flow is imposed by ejecting airflows from the four lateral wind walls.

It should be noted that the applicability of the computational approaches to the present problem relies on the two essential features of the small-scale flickering buoyant diffusion flames concerned. First, the flicker of the flames is due to the evolution of toroidal vortices, and the flames are away from the state of extinction followed by re-ignition. In the scenario, the characteristic time scales of chemical reactions are significantly smaller than those of the flow. Consequently, a complex reaction mechanism is not indispensable for computationally reproducing the flame flicker, and a simplified mixing-limited chemical reaction mechanism is adequate instead. Second, the flames concerned are of relatively small sizes, and the flow remains laminar in almost everywhere except in the far downstream of the flame, where the transitional or turbulent flow characteristics have however little influence on the vortices and the flame in the upstream. Consequently, more sophisticated computational models for turbulent flows, for example the large eddy simulation (LES), are not necessary in the present problem. The present computational results, to be shown shortly in the next sections, substantiate the reliability of the adopted computational methodology and models for small-scale flickering diffusion flames in weakly rotatory flows. For those flames of large scales or in strongly rotatory flows, the LES approach with finite-rate chemical reaction modelling is needed.

2.2 Computational Setup of Rotatory Flows

In order to computationally setup the externally rotatory flow, four lateral wind walls are loaded to eject horizontal airflows, as shown in Fig. 1(b). The inlet velocity on the wind wall is $\boldsymbol{U} = \boldsymbol{U}_\perp + \boldsymbol{U}_\parallel$, where $\boldsymbol{U}_\perp$ and $\boldsymbol{U}_\parallel$ are the normal and azimuthal velocity components respectively and the angle between them is fixed at $\alpha = 45°$. Consequently, the inlet airflows form a rotatory flow in the central region and the rotation intensity can be controlled by adjusting the magnitude of inlet airflow $U = |\boldsymbol{U}|$.

To facilitate the following presentation and discussion of results, $D$, $\sqrt{gD}$, and $\rho_\infty$ are used to nondimensionalize all kinematic and dynamic flow quantities. As a result, the dimensionless Cartesian coordinates are $(\hat{x}, \hat{y}, \hat{z}) = (x, y, z)/D$, the dimensionless time is $\hat{t} = t\sqrt{g/D}$, the dimensionless velocity components are $(\hat{u}_x, \hat{u}_y, \hat{u}_z) = (u_x, u_y, u_z)/\sqrt{gD}$, the dimensionless vorticity components are $(\hat{\omega}_x, \hat{\omega}_y, \hat{\omega}_z) = (\omega_x, \omega_y, \omega_z)\sqrt{D/g}$. To measure the intensity of the rotatory flow, we introduced the dimensionless parameter $R = \widehat{U}/\widehat{U}_0 = U/U_0$ and call it the rotational intensity number hereinafter. In addition, the coordinate transformation of $\hat{r} = \sqrt{\hat{x}^2 + \hat{y}^2}$ and $\hat{\theta} = \tan^{-1}(\hat{y}/\hat{x})/\pi$ is used for the cylindrical polar coordinates of $(\hat{r}, \hat{\theta}, \hat{z})$, where the velocity $(\hat{u}_r, \hat{u}_\theta, \hat{u}_z)$ and the vorticity $(\hat{\omega}_r, \hat{\omega}_\theta, \hat{\omega}_z)$ are calculated correspondingly [41].

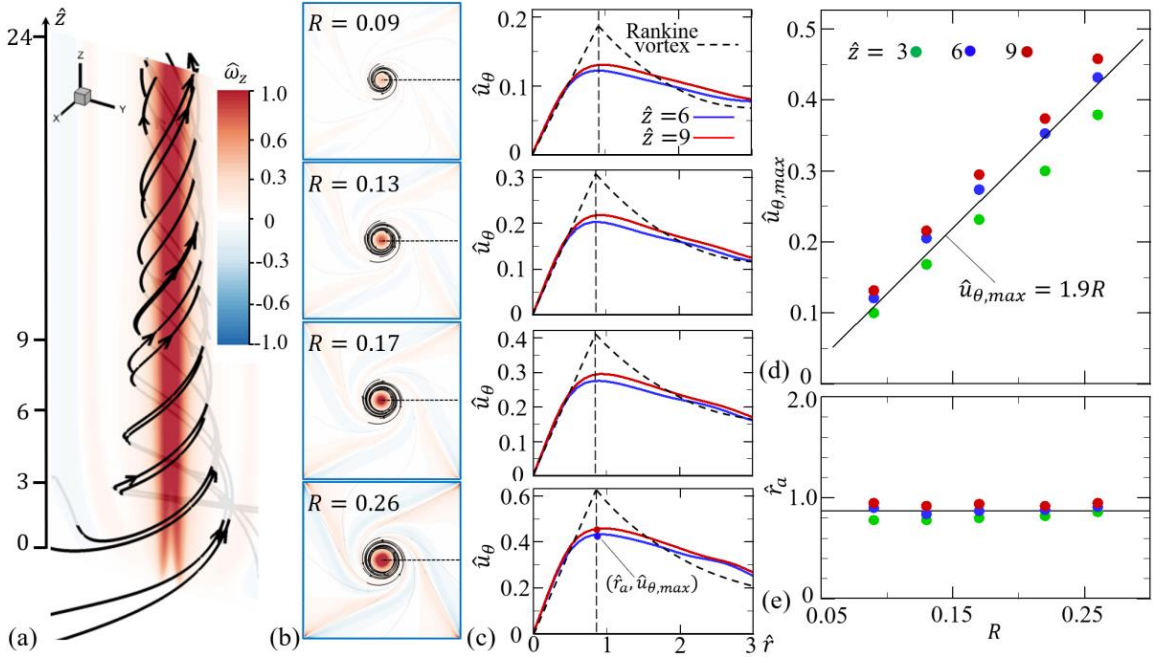

FIG. 2. The vertical component $\hat{\omega}_z$ of vorticity of a non-reactive methane jet in rotatory flows: (a) longitudinal section of $R = 0.17$ and (b) horizontal section at $\hat{z} = 9$. The four cases correspond to $R = 0.09$, 0.13, 0.17, and 0.26 respectively. (c) The radial profiles of azimuthal velocity $u_\theta$ in the four cases. The blue and red solid lines denote the vertical positions of $\hat{z} = 6$ and 9, respectively. The corresponding Rankine vortex approximation is plotted in the dashed line. (d) The maximum azimuthal velocity $\hat{u}_{\theta,max}$ and (e) the radial location $\hat{r}_a$ of vortex cores generated at different $R$.

To examine the generated rotatory flows, we simulated a few testing cases of non-reacting flows with different $R$, as shown in Fig. 2. The defined rotational intensity number $R$ varying from 0.09 to 0.26 is significantly smaller than unity, which justifies the "weakness" of the rotatory flows concerned. As shown in Fig. 2(a), the rotatory flow resembles a quasi-cylinder vortex, where the range and intensity of the vertical component $\hat{\omega}_z$ of vorticity vector remains approximately unchanged above $\hat{z} = 3$. Fig. 2(b) shows that the magnitude of $\hat{\omega}_z$ of the generated vortical flows increases with $R$, but the size of vortex core (illustrated by the streamline-encompassed area) keeps nearly the same. To quantify the vortical flow fields, we plotted the radial profiles of azimuthal velocity $\hat{u}_\theta$ for four cases in Fig. 2(c). It is seen that $\hat{u}_\theta$ linearly increases up to a maximum value (referred to as $\hat{u}_{\theta,max}$) at $\hat{r}_a$ and then gradually decays along the radial direction, which is similar to the experimentally measured transverse flow before the vortex breakdown occurs [31].

To facilitate the following comparison of computational and theoretical results, we approximated the calculated vortical flow as the Rankine vortex [41], which has an azimuthal velocity profile as $\hat{u}_\theta(\hat{r}) \sim \hat{r}$ within a vortex core of radius $\hat{r}_a$ and $\hat{u}_\theta(\hat{r}) \sim 1/\hat{r}$ outside the vortex core. As shown in Fig. 2(d), the present results show that $\hat{u}_\theta(\hat{r}_a) = \hat{u}_{\theta,max}$ is linearly proportional to $R$ with the proportionality being about 1.9. In addition, Fig. 2(e) shows that $\hat{r}_a$ is nearly a constant about 0.9, which indicates that the flickering flame is almost completely resided inside the vortex core.

2.3 Computational Validation

To validate the present computational methodology and models for capturing the flickering phenomenon of buoyant diffusion flames, we conducted a number of simulation runs with different $D$ and $g$ in quiescent environment (i.e., $R = 0$). For example, a snapshot of the flickering buoyant diffusion flame at $D = 10$ mm and $g = 9.8$ m/s$^2$ is illustrated in Fig. 3(a). Around the flame, the outside shear layer (denoted by the vorticity contour) rolls up to form the toroidal vortex (denoted by the curly streamlines). The flame is pinched off into two parts (the main flame anchored to the ground and the separated flame bubble) by the vortex at $\hat{z} = 6$. The dynamical process of flame flicker will be expounded in the next section.

In Fig. 3(b), the present results clearly show that the calculated frequencies of flickering flames agree well with previous experiments [6, 42-44] and particularly predict the famous scaling relation of $f_0 \sim \sqrt{g/D}$ [4, 11], where the proportionality factor 0.4 is close to 0.48 obtained from plenty experimental data reported in [11]. In addition, we found that the gravity changes of $(0.5g - 1.5g)$ can significantly affect the flickering frequency, confirming that the buoyancy is predominant in the present diffusion flames. The boundary flow near the bottom wall has negligible influence on the flickering of laminar diffusion flames, as the growth and shedding of the toroidal vortex occurs downstream far from the bottom. Carpio et al. [45] reported that, in a strongly rotatory flow, the bottom corner flow may result in the lift-off of flame base, which is absent in the present problem focusing on weakly rotatory flows.

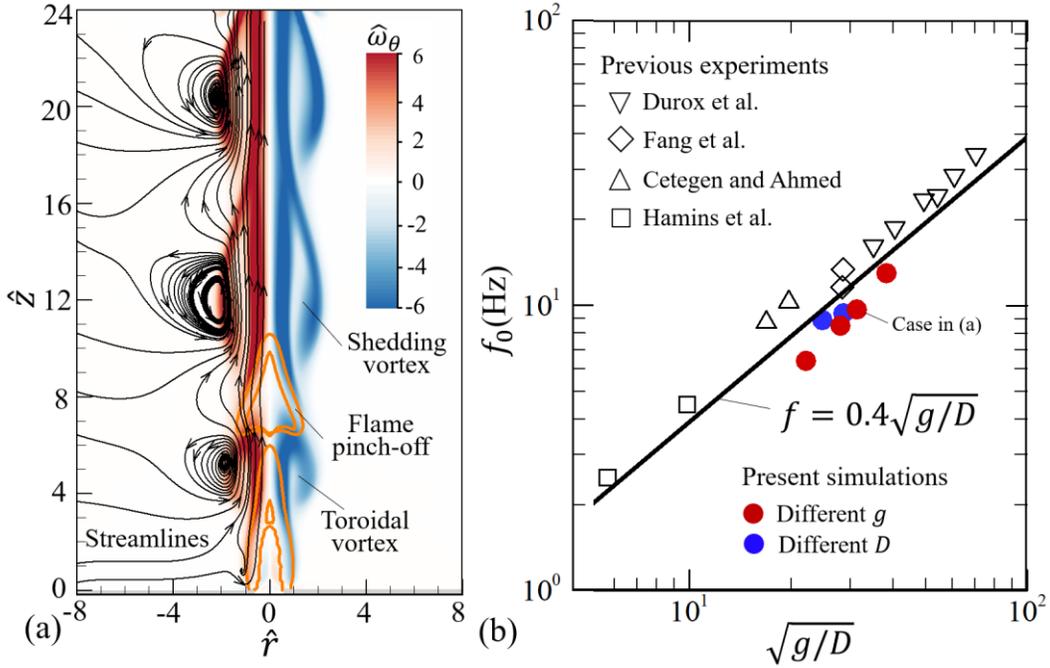

FIG. 3. (a) The streamlines and the contour of vorticity $\hat{\omega}_\theta$ of a flickering buoyant diffusion flame. The flame is represented by the orange isoline of heat release. (b) The validation of computational methods for single flickering jet flames.

### 3. Results and Discussion

In this section, we will answer the three questions raising in the Introduction. Particularly, we will clarify the influences of the externally rotatory flow on the buoyancy-induced toroidal vortex and on the variation of the flickering frequency and will theoretically establish the underlying mechanism of flickering flames in weakly rotational flows.

<u>3.1 Phenomenological Description</u>

In flickering buoyant diffusion flames, the buoyancy-induced shearing between the flame and the surrounding air is the precursor of the toroidal vortex [11, 18, 22]. During the lifecycle of the vortex, its formation, growth, and shedding correspond to the stretching, necking, and pinch-off of the flame, respectively. It can be seen in Fig. 4(a) for the case of $R = 0$ (without externally rotatory flow) that the vorticity layer forms at the flame base to stretch the flame (the normalized time $t^* = \hat{t}\hat{f}_0 = 0\sim0.3$), curlily grows along and necks the flame ($t^* = 0.3\sim0.8$), and sheds off ($t^* = 0.8\sim1.0$) to pinch off the flame. In this way, the flame performs a periodic flickering process. During $t^* = 1.0$~$1.1$, the shedding vortex moves downstream and the carried flame bubble burns out soon, while a new vortex generates at the flame base and a new cycle starts.

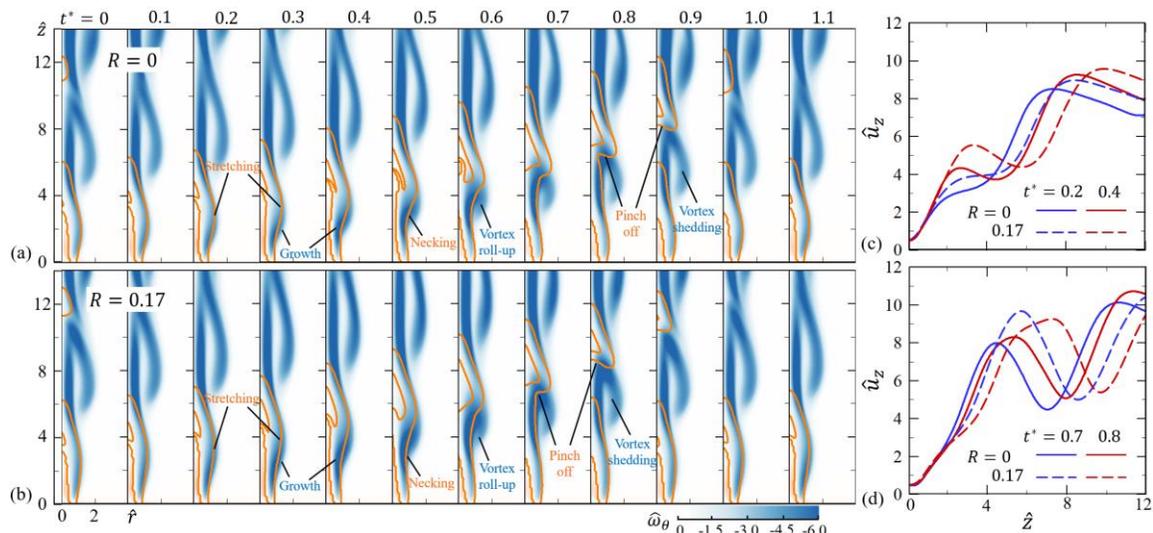

FIG. 4. The time-varying evolution of a flickering buoyant diffusion flame in (a) quiescent environment ($R = 0$) and (b) rotatory flow ($R = 0.17$). The comparison of their vertical velocity $\hat{u}_z$ along the central axis at (c) the instants $t^* = 0.2$ and 0.4 and (d) the instants $t^* = 0.7$ and 0.8.

Similar evolution of the toroidal vortex can also be observed in Fig. 4(b) for the typical case of $R = 0.17$, where the stretching, necking, and pinch-off of the flame remain in close associations with the formation, growth, and shedding of the vortex. An interesting observation is that the flame tends to be pinched off at a further downstream location with increasing $R$. Specifically, the flame pinch-off occurs at $\hat{z} = 6.0$ for the case of $R = 0$ but at $\hat{z} = 7.2$ for the case of $R = 0.17$. This observation can be understood by that the externally rotatory flow induces an additional vertical flow, which convects the vortex to the downstream. To substantiate the understanding, the central axial profiles of the vertical velocity $\hat{u}_z$ in the two cases are plotted in Fig. 4(c) and 4(d), respectively. It is seen that $\hat{u}_z$ rapidly increases from the fuel inlet and reaches the first peak at the position of the toroidal vortex. The comparison of $\hat{u}_z$ clearly indicates that the external rotation enhances the vertical flow (with higher peak values) and thus contributes to the additional motion of vortex (more downstream position of the first peak). The most interesting observation is that the flame pinch-off tends to occur earlier with increasing $R$. Specifically, the flame is pinched off at $t^* = 0.75$ for the case $R = 0.17$ compared with that at $t^* = 0.85$ for the case of $R = 0$. To further quantify and interpret this observation will be the focus of the following subsections.

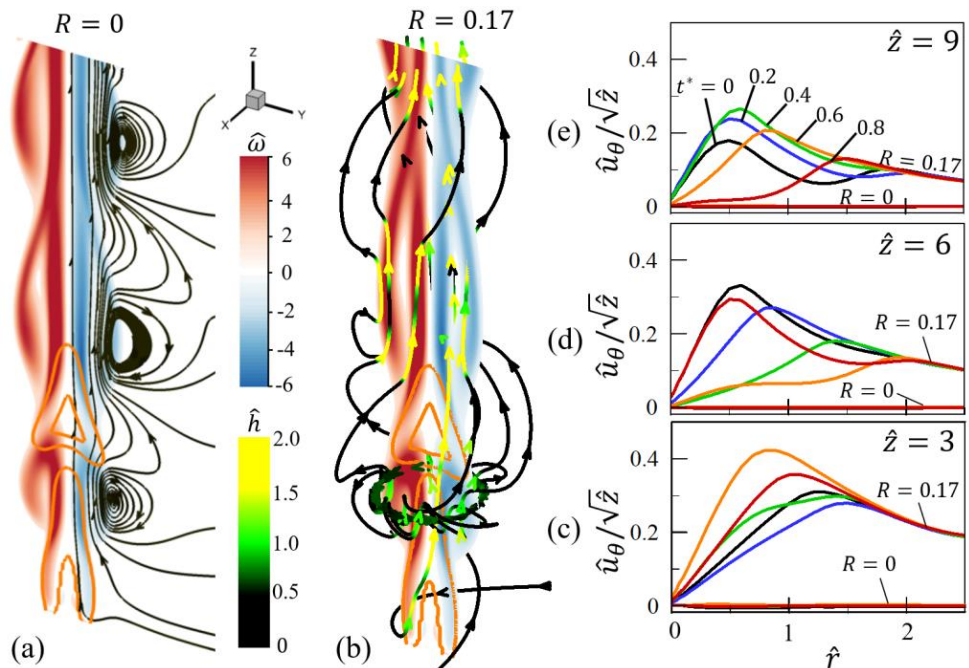

FIG. 5. Flickering buoyant diffusion flames in (a) quiescent environment and (b) rotatory flow, corresponding to the $R = 0$ and $R = 0.17$ cases respectively. The flame and vorticity $\hat{\omega}$ in the Y-Z plane are plotted. The helicity density $\hat{h}$ is dyed along the streamlines. The arrow on the streamlines denotes the flow direction. Their time-varying radial profiles of azimuthal velocity $\hat{u}_\theta$ at (c) $\hat{z} = 3$, (d) $\hat{z} = 6$, and (e) $\hat{z} = 9$.

The present results indicate that the dynamics of toroidal vortex is affected due to the rotation of ambient flow. To clearly show the effects of rotatory flow, we plotted streamlines around the flames and colored the local helicity density $\hat{h} = \hat{\boldsymbol{u}} \cdot \hat{\boldsymbol{\omega}}$ along the streamlines in Fig. 5(a) and 5(b) for the cases of $R = 0$ and $R = 0.17$ respectively. The zero value of $\hat{h}$ means that streamline is locally orthogonal to vorticity line, while the nonzero $\hat{h}$ can be used to quantify the local geometrical helix. Specifically, it can be seen in Fig. 5(a) that the flow around the flickering flame in quiescent environment is lamellar, because $\hat{h}$ is zero everywhere and the streamlines is restricted in the Y-Z plane due to the axis-symmetry. As the ambient flow is rotating, the lamellar flow is twisted toward the circumferential direction so that the streamlines tilt out of the Y-Z plane for forming a spiral ring, as shown in Fig. 5(b). Nonzero $\hat{h}$ appears in the vorticity layer around the flame and increases with the rotational intensity number $R$. Regardless of the local helix of the flow field, the flame morphology retains the approximate axis-symmetry within a certain extent. Therefore, in the present problem concerning weakly rotatory flows, there is only moderate symmetry breaking to the flame shape and surrounding shear layer, which will be used as a useful approximation to simplify our theoretical modelling to be expounded in the next section.

To estimate the extent to which the presence of flame affects the rotatory ambient flow, we replotted the radial profiles of azimuthal velocity $\hat{u}_\theta$ at different streamwise locations of $\hat{z} = 3$, 6, and 9. It is clearly seen that, although the profile of $\hat{u}_\theta$ changes with time due to the flame flicker, it retains the similarity to the Rankine vortex flow as $\hat{u}_\theta$ increases linearly with $\hat{r}$ within a certain range (i.e., the vortex core) that is sufficiently large to include the flame and its surrounding shear layer. It is also found that the decrease of $\hat{u}_\theta$ with $\hat{r}$ outside the vortex core is slower than the trend of $1/\hat{r}$ due to the additional flow induced by the thermal expansion around the flame. This does not have significant effects on the evolution of toroidal vortices within the vortex core. In addition, the decrease of $\hat{u}_\theta/\sqrt{\hat{z}}$ along the axial direction indicates that the circumferential motion caused by the externally rotatory flow becomes weaker at the downstream. Consequently, the buoyancy-induced vortex flow still plays the predominated role in the present flickering flames in weakly rotatory flows, which will facilitate the following modelling in the next section.

Previous study [11] has shown that the pressure gradient $\nabla p$ can be negligible for the vorticity generation in flickering flames if the ambient flow is quiescent and the Froude number is small. In the present problem, we hypothesized that the formation and evolution of toroidal vortex around the flame is affected by the baroclinity $\nabla \rho \times \nabla p$ due to the externally rotatory flow. To verify this hypothesis, we plotted the radial pressure gradient $\partial \hat{p}/\partial \hat{r}$ around the flames in the cases of $R = 0$ and $R = 0.17$, as shown in Fig. 6(a). It is seen that the magnitude of $\partial \hat{p}/\partial \hat{r}$ is not negligible in almost entire flow domain and is significant in the domain close to the flame. Besides, Fig. 6(b) shows that the azimuthal pressure gradient $(\partial \hat{p}/\partial \hat{\theta})/\hat{r}$ has much smaller magnitude than $\partial \hat{p}/\partial \hat{r}$ so that we can neglect the azimuthal component in the following analysis. Consequently, the pressure gradient along the radial direction should be considered in the present problem concerning the influence of the rotatory flows.

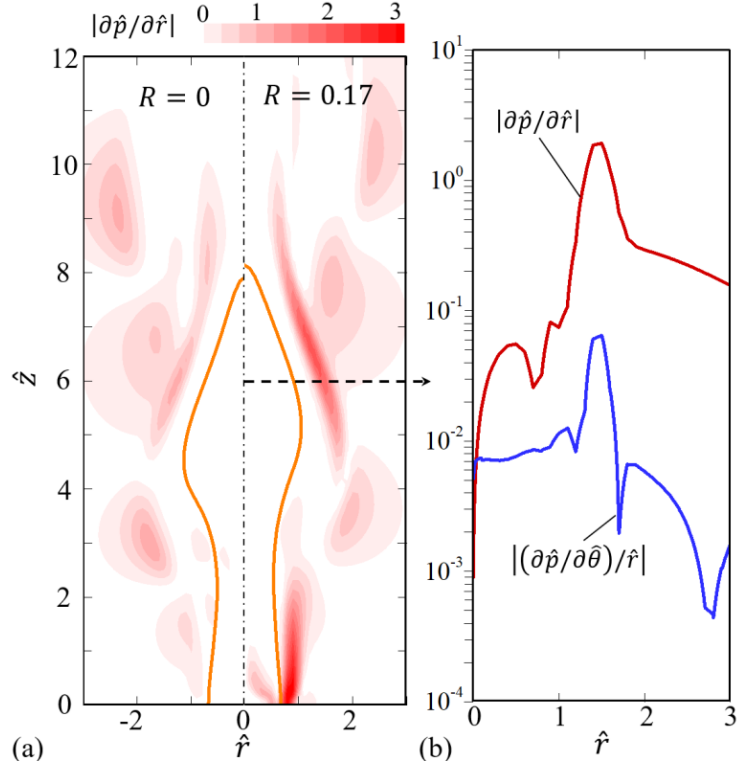

FIG. 6. (a) The radial component of pressure gradient $|\partial \hat{p}/\partial \hat{r}|$ for the cases of $R = 0$ and $R = 0.17$, corresponding to the left and right subfigures respectively. (b) The radial component $|\partial \hat{p}/\partial \hat{r}|$ and azimuthal component $|(\partial \hat{p}/\partial \hat{\theta})/\hat{r}|$ of pressure gradient at $\hat{z} = 6$ in the case of $R = 0.17$.

3.2 Influence of $R$ on Flickering Frequency

To determine the frequency of flame flicker, it is required to choose an appropriate quantity that can characterize the dynamical behavior of flickering flames. In the previous experiments, there is quite freedom in selecting and acquiring characteristic (either local or global) flame quantities, for example, the pressure or temperature of flickering flame at a certain position [6, 46]; the flame luminosity at a certain height [47, 48] ; the flame morphology information (e.g., the flame height [1], width [27], size [49] and brightness [13, 21]) obtained from high-speed images. In the present computational work, we choose the total heat release rate $Q$ and the vertical velocity $u_z$ and temperature $T$ at a fixed point to determine the frequency of flame flicker, as shown in Fig. 7(a), 7(b) and 7(c) respectively. The former one can be treated as a representative global quantity and the latter two are representative local quantities. The present results show that all the quantities exhibit periodic behaviors with almost the same frequency for the ambient flow being rotatory or not, while there are slight phase differences among them. This result is consistent with previous experimental observations [6, 21] that the dominant frequencies based on different quantities in the flickering flame have negligible difference. For simplicity and consistency, the global quantity $Q$ will be adopted in the present study for analyzing the frequency variation in different cases.

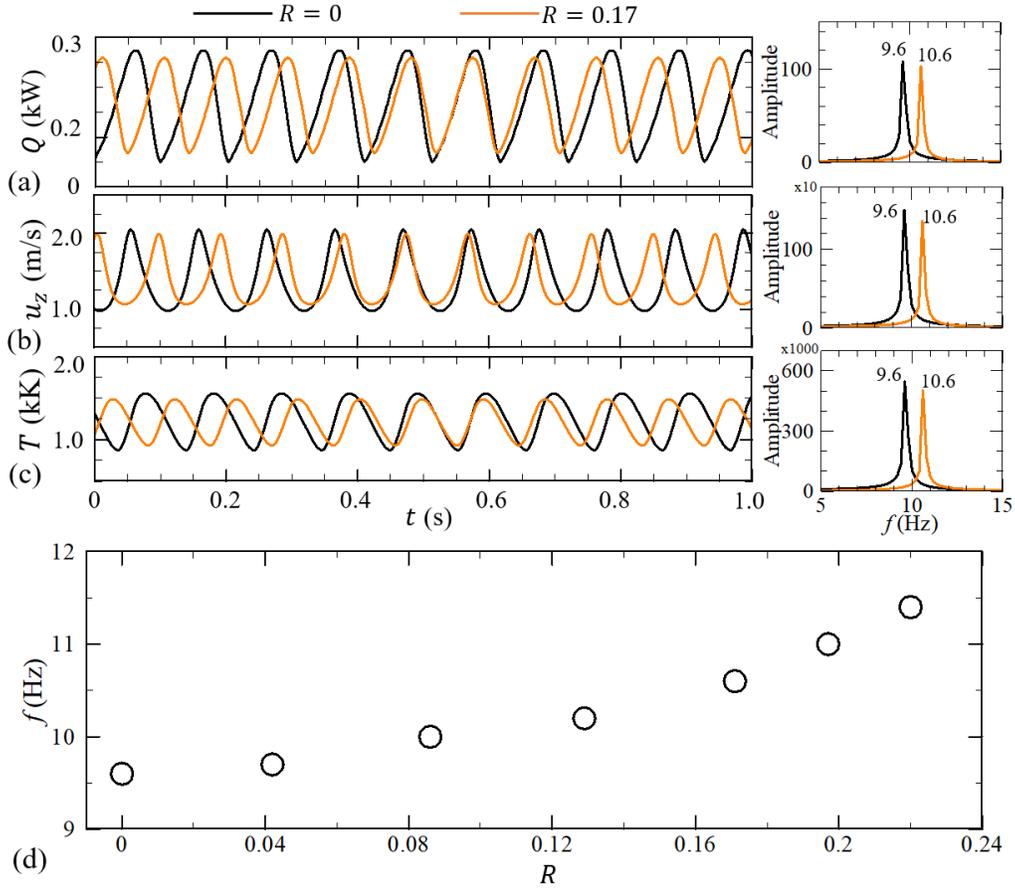

FIG. 7. Time and frequency domain graphs of (a) the total heat release rate $Q$, (b) the vertical velocity $u_z$ and (c) the temperature $T$ at $\hat{z} = 3$ of the central axis for flickering buoyant diffusion flames. The black and orange lines correspond to the $R = 0$ and $R = 0.17$ cases, respectively. (d) The correlation between the flickering frequency $f$ and the rotatory number $R$.

It is seen in Fig. 7(a) that the periodic wave of $Q$ for the case of $R = 0$ oscillates at the frequency of $f_0 = 9.6$ Hz, which is smaller than 10.6 Hz for the case of $R = 0.17$. The frequency comparison further confirms the observation in Sec. 3.1 that the externally rotatory flow causes the flame pinched off at an early time. Fig. 7(d) shows a monotonically increasing trend of the frequency $f$ with the rotatory number $R$, and the trend is noticeably nonlinear. It should be noted that the present study focuses on the weakly rotatory flows with $R < 0.24$, beyond which the stronger rotational flow may lead to the occurrence of vortex breakdown and local flame extinction [31, 45, 50]. More sophisticated computational approaches are needed to deal with these emerged phenomena in strongly rotatory flows and merit future work.

3.3 Theoretical Model of Flickering Flame in a Weakly Rotatory Flow

The above computational results hitherto have demonstrated the influence of weakly rotatory flows on the formation of toroidal vortex and the frequency augmentation. Next, we will reveal the underlying mechanism. Following the theory of Xia and Zhang [11], the present theoretical model for flame flicker is based on its connection with the shed-off of the toroidal vortex. To facilitate the comparison with the theory of Xia and Zhang [11], we adopted the same notation within this subsection. The central idea of the theory is that the flame flicker occurs when the circulation of the toroidal vortex reaches a threshold, which is an approximately universal constant [51-54] and presumably does not have significant change in the present weakly rotatory environments. The frequency relation for buoyancy-driven diffusion flames in weakly rotatory flows will be theoretically derived through three steps as follows.

In the first step, the generation rate of total circulation $\Gamma$ inside a control mass $A$ is formulated based on the approximation of axis-symmetry, as discussed in Section 3.1. As illustrated in Fig. 8(a), a vortex layer segment around the flame is encircled by the red dashed box, which is the material contour $\partial A$ of a control mass $A$. The zoomed subfigure clearly shows that the domain of $A$ is vertically between $S_{v1}$ and $S_{v2}$ and radially between $S_{r1}$ and $S_{r2}$, where a material line element is represented as $\boldsymbol{s}ds$ and $\boldsymbol{s}$ is the unit tangential vector along the contour $\partial A$. According to the Kelvin's circulation formula [55], the rate of total circulation change is

$$\dot{\Gamma} = \oint_{\partial A} \boldsymbol{a} \cdot \boldsymbol{s} ds \qquad (1)$$

where $\dot{\Gamma} = d\Gamma/dt$ is the change rate of circulation, $\boldsymbol{a} = D\boldsymbol{u}/Dt = -[\nabla p + (\rho_\infty - \rho)\boldsymbol{g}]/\rho + \nu \nabla^2 \boldsymbol{u}$ is the acceleration of control mass [11]. As the diffusion term is not a source of vorticity production and vanishes on $\partial A$, the dimensionless $\hat{\boldsymbol{a}} = (D\boldsymbol{u}/Dt)/g$ can be expressed as

$$\hat{\boldsymbol{a}} = -\frac{\nabla \hat{p}}{\hat{\rho}} + \left(1 - \frac{1}{\hat{\rho}}\right)\hat{\boldsymbol{g}} \quad (2)$$

For the pressure gradient term $\nabla \hat{p}$, we only consider the radial component $\partial \hat{p}/\partial \hat{r}$ as discussed in Sec. 3.1.

Considering that the present rotatory flows resemble the Rankine vortex, we use the relationship between the azimuthal velocity and the radial pressure gradient

$$\frac{\partial \hat{p}}{\partial \hat{r}} = \frac{\hat{\rho}\hat{u}_\theta^2}{\hat{r}} \quad (3)$$

where the azimuthal velocity $\hat{u}_\theta$ has a linear correlation with $\hat{r}$, namely $\hat{u}_\theta = C_\theta \hat{r}$ with a measurable quantity $C_\theta = \hat{u}_\theta(\hat{r}_a)/\hat{r}_a$ for a given rotatory flow. Applying Eq. (3) and Eq. (2) in Eq. (1), we have

$$\hat{\Gamma} = \oint_{\partial A} -C_\theta^2 \hat{r} \boldsymbol{r} \cdot \boldsymbol{s} d\hat{s} + \oint_{\partial A} (1 - \frac{1}{\hat{\rho}})\hat{\boldsymbol{g}} \boldsymbol{z} \cdot \boldsymbol{s} d\hat{s} \quad (4)$$

where $\boldsymbol{r}$ is the unit radial vector and $\boldsymbol{z}$ is the normal vector. The first term is the radial integration along $\partial A$ due to the externally rotatory flow, while the second term is the vertical integration along $\partial A$ due to the buoyance-induced flow. Compared with the theory of Xia and Zhang [11] for flickering flame in a quiescent environment, the first term in Eq. (4) is due to the baroclinic effect caused by the rotatory flow. The two terms can be further simplified as

$$\oint_{\partial A} -C_\theta^2 \hat{r} \boldsymbol{r} \cdot \boldsymbol{s} d\hat{s} = -C_\theta^2 \int_{S_{r1}}^{S_{r2}} d\hat{r}^2 \quad (5a)$$

$$\oint_{\partial A} (1 - \frac{1}{\hat{\rho}})\hat{\boldsymbol{g}} \boldsymbol{z} \cdot \boldsymbol{s} d\hat{s} = \left(\frac{1}{\hat{\rho}_f} - \frac{1}{\hat{\rho}_\infty}\right) \int_{S_{v1}}^{S_{v2}} \hat{g} d\hat{z} \quad (5b)$$

where $\hat{\rho}_f = \rho_f/\rho_\infty$ and $\hat{\rho}_\infty = 1$ are the dimensionless density inside and outside the vortex layer respectively. Integrating along the paths of $S_{r1} - S_{r2}$ and $S_{v1} - S_{v2}$, as shown in Fig. 8(a), we can rewrite Eq. (4) as

$$\hat{\Gamma} = -[2C_\theta^2 \hat{r}_c \Delta \hat{r} + (C_\rho - 1)\hat{g}\Delta \hat{z}] \quad (6)$$

where $\hat{r}_c$ is the radial position of the vortex layer and close to the flame sheet; the density ratio $C_\rho = \rho_\infty/\rho_f$ is a measurable quantity for a given flame, for example $C_\rho$ is about 7.5 for the present computational methane/air flames, $\Delta \hat{r}$ and $\Delta \hat{z}$ are the radial and vertical unit lengths of the vortex layer associated with the control mass $A$. It should be noted that the first term of Eq. (6) is attributed to the externally rotatory flow, which is absent in the theory of Xia and Zhang [11].

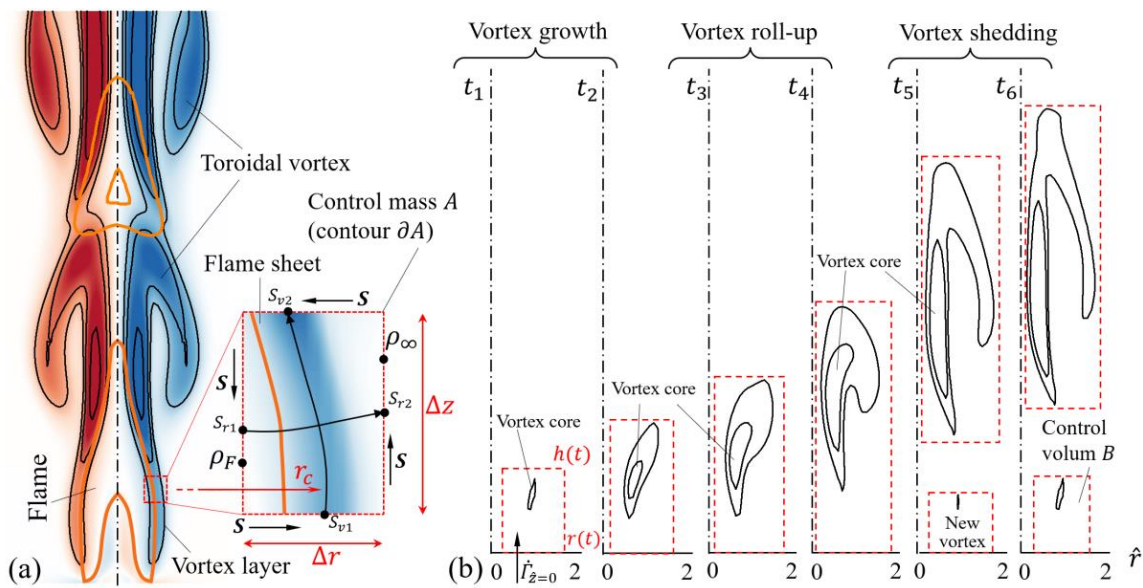

FIG. 8. (a) The vortex layer around flame and the exaggerated segment for illustrating the control mass $A$. (b) The periodic formation process of a toroidal vortex, which is represented by the vorticity contours and tracked by the red dashed box for the control volume $B$.

In the second step, the periodic formation process of a toroidal vortex associated with the flame flicker is established. Fig. 8(b) illustrates the evolution of a toroidal vortex in the moving control volume $B$. During $t_1 \sim t_2$, a new vortex core of the toroidal vortex generates near the base of the flame and then grows downstream under the buoyancy-induced convection. During $t_3 \sim t_4$, the vortex head region rolls up outward and the central vortex core moves downstream. During $t_5 \sim t_6$, the vortex fully develops and detaches. Meanwhile a new vortex is formed at the flame base and a new cycle starts. It can be seen in Fig. 8(b) that the increment of $\Delta \hat{z}$ is much larger than that of $\Delta \hat{r}$ during a periodic motion and $\hat{r}_c$ of the vortex core is approximately a constant. For the total circulation $\hat{\Gamma}_B$ of the moving vortex enclosed in $B$, its change rate should include an additional term $\dot{\hat{\Gamma}}_{\hat{z}=0} = -C_j \hat{U}_0^2$ [11, 56], which is caused by the fuel inlet. According to Eq. (6), then we have

$$\hat{\Gamma}_B(\hat{t}) = -2C_\theta^2 \hat{r}_c \hat{r}(\hat{t}) - (C_\rho - 1)\hat{g}\hat{h}(\hat{t}) - C_j \widehat{U}_0^2 \tag{7}$$

where $\hat{r}(\hat{t})$ is the width of the right boundary of $B$, $\hat{h}(\hat{t})$ is the height of the upper boundary of $B$, and $-C_j \widehat{U}_0^2$ enters the system through the lower boundary of $B$, where the constant $C_j$ relies on the configuration and the jet inlet condition.

Focusing on the formation of the toroidal vortex from the new to the fully developed, we integrate Eq. (7) in a period $\hat{t} = U_0 \tau / D$ and have

$$\hat{\Gamma}_{TV} = \int_0^{\hat{t}} \hat{\Gamma}_B d\hat{t} = -2C_\theta^2 \hat{r}_c \overline{R}\hat{t} - (C_\rho - 1)\hat{g}\overline{H}\hat{t} - C_j \widehat{U}_0^2 \hat{t} \tag{8}$$

where $\overline{R} = \hat{t}^{-1} \int_0^{\hat{t}} \hat{r}(t) dt$ and $\overline{H} = \hat{t}^{-1} \int_0^{\hat{t}} \hat{h}(t) dt$ represent the time-averaged width and height of growing toroidal vortex, respectively. According to the present computational results that the width of toroidal vortex is nearly unchanged and the vortex core moves close to the flame sheet, we have $\overline{R} \approx 1.9$ and $\overline{\hat{r}_c} \approx 0.6$. In addition, $h(t)$ can be roughly estimated as $\sqrt{gD}t$, as the toroidal vortex is driven downstream by the buoyancy. Thus, we have $\overline{H} = C_h \hat{t} / \widehat{U}_0$, where $C_h$ is a constant prefactor. With being scaled by $-\widehat{U}_0$, Eq. (8) can be rewritten as

$$\hat{\Gamma}_{TV}^* = C_h Ri \hat{t}^2 + (C_j + C_r)\sqrt{Fr}\hat{t} \tag{9}$$

where $Ri = (C_\rho - 1)\hat{g}/\widehat{U}_0^2 = (C_\rho - 1)gD/U_0^2$ is the Richardson number, $Fr = \widehat{U}_0^2 = U_0^2/gD$ is the Froude number, and $C_r = 2C_\theta^2 \hat{r}_c \overline{R}/\widehat{U}_0^2$ is a prefactor for characterizing the externally rotational flow.

In the third step, a threshold for the accumulation of the circulation inside the toroidal vortex [51, 54] is applied for obtaining the frequency relation. Applying $\hat{\Gamma}_{TV}^* = C$ to Eq. (9), we have

$$\frac{f}{\sqrt{g/D}} = \frac{\widehat{U}_0}{\hat{t}} = \frac{1}{2C}\left(C_{jr} Fr + \sqrt{C_{jr}^2 Fr^2 + CC_h C_\rho}\right) \tag{10}$$

where $C_{jr} = C_j + C_r$ is a prefactor for the combined contributions of the initial jet flow and the externally rotatory flow. Hereto, we complete the derivation of the frequency relation for flickering buoyant diffusion flames in rotatory flows. For the case of $R = 0$, the prefactor $C_{jr}$ degenerates to $C_j$, and the frequency of flickering flames in quiescent environment is obtained as

$$\frac{f_0}{\sqrt{g/D}} = \frac{1}{2C}\left(C_j Fr + \sqrt{C_j^2 Fr^2 + CC_h C_\rho}\right) \qquad (11)$$

which is exactly the same with the scaling formula obtained by Xia and Zhang [11]. It is interesting to see that Eq. (10) and Eq. (11) have the same functional form and differ in only the prefactor of the $Fr$-term. The underlying physics is that the externally rotatory flow plays a similar role as that of the initial jet for enhancing the vortex growth.

3.4 Comparison between Computation and Theory.

Next, we compare the above theoretical formula with the present computational results. Based on Eq. (10), we can have the frequency increase of flickering buoyant diffusion flames due to the additional rotation of the ambient flow

$$\Delta \hat{f} = \hat{f}(R) - \hat{f}(R=0) = \frac{f(R) - f(R=0)}{\sqrt{g/D}} = \frac{5\Phi}{C} R^2 \qquad (12)$$

where $\Phi = 1 + (C_j + C_{jr})/\left(\sqrt{C_j^2 + CC_h C_\rho/Fr^2} + \sqrt{C_{jr}^2 + CC_h C_\rho/Fr^2}\right)$ is a constant factor about 1~2, and the approximation of $C_r = 10R^2/\widehat{U}_0^2$ is used due to $C_\theta = 2.11R$ and $\hat{r}_c \overline{R} = 1.14$. Therefore, we obtain a scaling law of $\Delta \hat{f} \sim R^2$ from Eq. (12). As shown in Fig. 9, the scaling law agrees very well with the present computational results.

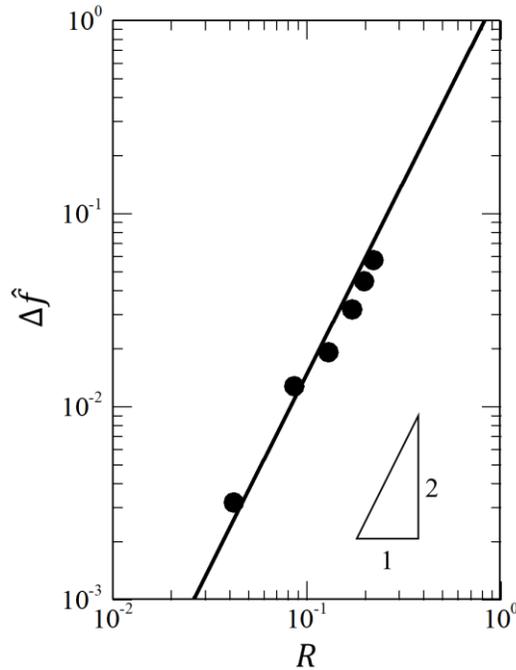

FIG. 9. Comparison between the correlation of $\Delta \hat{f} \sim R^2$ with the data in Fig. 7(d).

It should be emphasized that the present theoretical analysis is focused on the weakly rotatory flow. Consequently, the $\Delta \hat{f} \sim R^2$ scaling law is limited to sufficiently small $R$. When $R$ is larger than 0.24 in the present study, the flame flicker may not be clearly observed, and the flame oscillates in its tip with a slight swing. This result implies that the circumferential motion of toroidal vortex is not negligible and the axis-symmetry approximation becomes questionable. In addition, the sufficiently large $R$ is likely to cause the occurrence of vortex breakdown and local flame extinction. This problem requires a completely differential computational and theoretical framework and merits further research.

## 4. Concluding Remarks

Flickering diffusion flames in quiescent environment have been extensively studied in the literature, but their characteristics in complex external flows, particularly in externally rotatory flows, was investigated inadequately. The present study presents computational and theoretical investigations on the small-scale flickering buoyant diffusion flames in weakly rotatory flows and the conclusions are summarized as follow.

First, four lateral forced-ventilation walls were computationally imposed to generate a rotatory flow with variable rotational speed $U$ around a methane jet diffusion flame with fixed jet velocity $U_0$ ($Re = 100$ and $Fr = 0.28$). The rotational intensity is controlled by the defined dimensionless number $R = U/U_0$. The generated rotatory flow was found to resemble the Rankine vortex as it has a linear velocity profile within a vortex core whose size slightly changes with $R$. As a validation, flickering buoyant diffusion flames of methane gas in quiescent environment were computationally reproduced and the calculated flicker frequencies $f_0$ agree well with the famous scaling relation $f_0 \sim \sqrt{g/D}$.

Second, the present computational results show that the externally rotatory flow enhances the periodic flickering motion and accord with the experimental observations reported in the literature. Furthermore, the flicker frequency $f$ is found to nonlinearly increases with $R$ up to 0.24. By analyzing the flow and pressure fields around the flame, we found that there is only slight axis-symmetry breaking to the flame shape and surrounding shear layer in weakly rotatory flows and that the radial pressure gradient is significantly increased compared with that at $R = 0$. The approximate axis-symmetry was subsequently used to simplify our theoretical investigation, and the significant radial pressure gradient implies that baroclinic effect must be taken into account in our theory.

Third, we formulated the scaling theory for interpreting the frequency increase of flickering buoyant diffusion flames in weakly rotatory flows. This theory can be degenerated to that of Xia and Zhang [11] at $R = 0$. The predicted frequency correlation of $\Delta \hat{f} = \hat{f} - \hat{f}_0 \sim R^2$ agrees very well with the present computational results. The underlying physics can be understood as that the externally rotatory flow enhances the pressure gradient in the radial direction, and the significant baroclinic effect $\nabla p \times \nabla \rho$ contributes an additional source for the growth of toroidal vortices. Consequently, the toroidal vortices reach the threshold of circulation for shedding in an early time.

It should be noted that the present study is concerned with the weakly rotatory flows, as a sufficiently large $R$ tends to cause the occurrence of vortex breakdown and even local flame extinction. The flame phenomena in strongly rotatory flows are govern by different physical mechanisms and require differential computational ant theoretical treatment, which are of interest for future work.

## Acknowledgements

This work is financially supported by the National Natural Science Foundation of China (No. 52176134) and by the Hong Kong Polytechnic University (G-UAHP).